# Precision spatial measurement of the hot rubidium atom in the three-dimension

Rahmatullah,[1,2] You-Lin Chuang,[2] Ray-Kuang Lee,[2] and Sajid Qamar[1(a)]

[1] *Quantum Optics Lab, Department of Physics, COMSATS institute of information technology, Islamabad, Pakistan*
[2] *Institute of Photonics and Technologies, National Tsing-Hua University, Hsinchu 300, Taiwan*

**Abstract** – The interaction of hot atoms with laser fields experiences a Doppler shift which can severely affect the precise spatial measurement of an atom. We suggest an experimentally realizable scheme to address this issue in three-dimension position measurement of a single atom in vapors of rubidium atoms. Three-level $\Lambda$ – type atom-field configuration is considered where a moving atom interacts with three orthogonal standing-wave laser fields and spatial information of the atom in 3D space is obtained via upper-level population using a weak probe laser field. The atom moves with a velocity 'v' along the probe laser field and due to the Doppler broadening the precision in the spatial information deteriorates significantly. It is reported that via a microwave field the precision in the position measurement of the single atom can be obtained in the hot rubidium atom overcoming the limitation posed by the Doppler shift.

**Introduction** – The quest to achieve precision in position measurement at atomic scale is not new. However, such kind of measurements is diffraction limited [1] and the spatial resolution can not be achieved better than the length scale given by the wavelength of the light field which is used for measurement [2]. Other than the fundamental nature of the problem, recent advances in laser cooling [3], lithography [4], Bose-Einstein condensate [5] and measurement of the centre-of-mass wave function [6-7] have made it more important to obtain precise position information of the atom. A lot of success has already been observed to achieve precision in spatial resolution with near- and far-field imaging techniques [8]. Further, some schemes have been proposed to obtain structures beyond the diffraction limit using position-dependent dark states for nano-scale resolution fluorescence microscopy [9] and in interferometric lithography [10].

In parallel, several proposals have been made to obtain position information of moving atom using quantum optical methods. In these proposals standing-wave driving fields have been used to encode the position information into the intensity pattern via position-dependent Rabi frequency. In the beginning, several schemes have been suggested for position measurement of the atom in one-dimension (1D) [11-16]. However, during the last few years, the interest has been shifted towards two-dimension (2D) atom localization and several schemes have been suggested in this regard [17-20]. It is clear that the next objective would be to obtain three-dimension (3D) atom localization.

Some progress has already been made and recently a scheme has been proposed for 3D atom localization in a four-level tripod type atom-field system using three orthogonal standing-wave fields [21] where eight possible positions of the atom within cubic optical wavelength in 3D space have been noticed. In a recent article, Wang and Yu have proposed 3D localization of cold $^{87}$Rb atom with 100% probability using probe absorption in a three-level atomic system [22].

A major limitation of almost all those schemes where the atom is considered moving through the fields that no Doppler shift has been incorporated. In fact, even for the static (cold) atom there is a chance that atom does not remain perfectly stationary when it is driven by the laser fields. In that case, the motion which could be modeled as a Gaussian velocity distribution, would affect the precise position measurement of the single atom due to the Doppler broadening.

In this letter, we consider hot rubidium ($^{87}$Rb) atom and propose a scheme for precision enhancement in position measurement of single atom in 3D space in the presence of Doppler broadening. Initially, we consider a three-level $\Lambda$-type atom-field configuration to obtain precise spatial information of an atom via upper-level population ignoring the Doppler shift which is in accordance with earlier work [22]. However, considering the more realistic experimental realization, we incorporate effect of Doppler broadening which reduces, significantly, the precision in position measurement of the single atom. To cater this problem, we apply an external microwave field between the two lower levels. The $\Lambda$-type atom-field configuration then becomes $\Delta$-type configuration and we observe that the effect of Doppler shift is significantly reduced via control of the microwave field. This implies that precision in the position measurement in 3D space can be retained even in the presence of Doppler shift.

**Model and equations** – In Fig. 1, we show the schematic of a three-level atom, having one upper-level ($|1\rangle$) and two lower-levels ($|2\rangle$ and $|3\rangle$). The atom interacts with a standing-wave field $E_{xyz}$, which is the superposition of three orthogonal standing-wave fields and couples the atomic transition

---

[(a)] E-mail: sajid_qamar@comsats.edu.pk

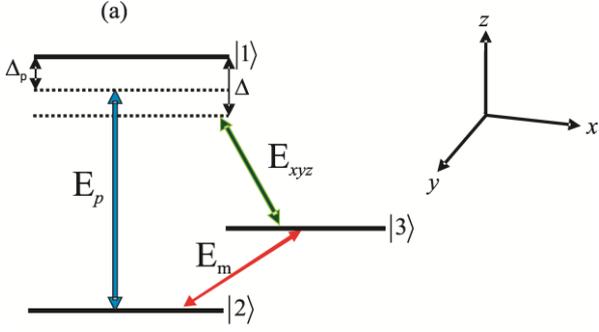

Fig. 1. The standing-wave field $E_{xyz}$ which is the superposition of three orthogonal standing-wave fields drives the atomic transition $|1\rangle \leftrightarrow |3\rangle$ and a weak probe field $E_p$ couples atomic transition $|1\rangle \leftrightarrow |2\rangle$. An additional microwave field $E_m$ is applied corresponding to the transition $|2\rangle \leftrightarrow |3\rangle$.

$|1\rangle \leftrightarrow |3\rangle$. To measure upper-level population, a weak probe laser field $E_p$ is applied which couples the atomic transition $|1\rangle \leftrightarrow |2\rangle$. The resulting atom-field configuration is $\Lambda$-type which is experimentally realizable assuming transition $5S_{1/2} \rightarrow 5P_{3/2}$ in $^{87}$Rb, the energy levels can be considered as $|1\rangle = |5P_{3/2}, F=2\rangle, |2\rangle = |5S_{1/2}, F=2\rangle$ and $|3\rangle = |5S_{1/2}, F=1\rangle$ [24] and has been used by Wang et. al for EIT [25] and for the 3D atom localization [22]. An additional microwave field $E_m$ is applied to the atomic transition $|2\rangle \leftrightarrow |3\rangle$ which controls the damaging effects of Doppler broadening. The atom-field configuration thus becomes $\Delta$-type, see Fig. 1. The corresponding Rabi frequencies are $\Omega_p$, $\Omega_{xyz}$ and $\Omega_m$, we consider $\Omega_{xyz}$ position-dependent.

The standing-wave field $E_{xyz}$ is considered as the superposition of three orthogonal standing-wave fields, i.e., $E_x$, $E_y$ and $E_z$ along the $x$, $y$ and $z$ directions, respectively. We also assume that each standing-wave field is also a superposition of the two standing-wave fields along the corresponding directions. We define position-dependent Rabi frequency as $\Omega_{xyz} = \Omega_x + \Omega_y + \Omega_z$, with

$$\Omega_x = \Omega_1 [\sin(k_1 x + \eta) + \sin(k_2 x)],$$
$$\Omega_y = \Omega_2 [\sin(k_3 y + \varsigma) + \sin(k_4 y)],$$
$$\Omega_z = \Omega_3 [\sin(k_5 z + \varphi) + \sin(k_6 z)], \quad (2)$$

where $k_i = 2\pi/\lambda_i (i=1,2,3,4,5,6)$ be the wave-vectors having wavelengths $\lambda_i (i=1,2,3,4,5,6)$ of the corresponding standing-wave fields. The parameters $\eta$, $\varsigma$ and $\varphi$ are the phase shifts associated with the standing-wave fields with wave-vectors $k_1$, $k_3$ and $k_5$, respectively.

The interaction picture Hamiltonian for the system in the dipole and rotating-wave approximation can be written as

$$H = -\hbar \left[ \Omega_p e^{i\Delta_p t} |1\rangle\langle 2| + \Omega_{xyz} e^{i\Delta t} |1\rangle\langle 3| + \Omega_m |3\rangle\langle 2| + H.c \right], \quad (3)$$

where detuning $\Delta_p = \omega_{12} - \nu_p$ and $\Delta = \omega_{13} - \nu$ are associated with the probe and standing-wave fields corresponding to the atomic transitions $|1\rangle \leftrightarrow |2\rangle$ and $|1\rangle \leftrightarrow |3\rangle$, respectively.

The equations of motion for the corresponding density matrix elements can be written as

$$\dot{\rho}_{11} = -(\Gamma_{12} + \Gamma_{13})\rho_{11} - i\Omega_p(\rho_{12} - \rho_{21}) - i\Omega_{xyz}(\rho_{13} - \rho_{31}),$$
$$\dot{\rho}_{22} = \Gamma_{12}\rho_{11} + \Gamma_{32}\rho_{33} + i\Omega_p(\rho_{12} - \rho_{21}) + i\Omega_m(\rho_{32} - \rho_{23}),$$
$$\dot{\rho}_{33} = \Gamma_{13}\rho_{11} - \Gamma_{32}\rho_{33} + i\Omega_{xyz}(\rho_{13} - \rho_{31}) - i\Omega_m(\rho_{32} - \rho_{23}),$$
$$\dot{\rho}_{23} = i(\Delta_p - \Delta + i\gamma_{23})\rho_{23} + i\Omega_p\rho_{13} - i\Omega_{xyz}\rho_{21} - i\Omega_m(\rho_{22} - \rho_{33}),$$
$$\dot{\rho}_{21} = i(\Delta_p + i\gamma_{21})\rho_{21} - i\Omega_p(\rho_{22} - \rho_{11}) - i\Omega_{xyz}\rho_{23} + i\Omega_m\rho_{13},$$
$$\dot{\rho}_{31} = i(\Delta_p + i\gamma_{31})\rho_{31} - i\Omega_{xyz}(\rho_{33} - \rho_{11}) - i\Omega_p\rho_{32} + i\Omega_m\rho_{12}, \quad (4)$$

here $\Gamma_{ij}(ij \in 1,2,3, i \neq j)$ is the decay rate from level $|i\rangle$ to level $|j\rangle$ whereas $\gamma_{23} = \Gamma_{32}/2$, $\gamma_{21} = (\Gamma_{12} + \Gamma_{13})/2$ and $\gamma_{31} = (\Gamma_{32} + \Gamma_{12} + \Gamma_{13})/2$.

Our objective is to obtain position information of the atom via absorption of probe field which is directly related to the upper-level population $\rho_{11}$. Initially, the atom is considered in ground state $|2\rangle$ and when the probe field is absorbed the atom is excited to the upper-level $|1\rangle$. The probability of upper-level population can thus be termed as the conditional position probability of the atom in 3D space.

To understand the dependence of the upper-level population on certain parameters like $\Omega_p$, $\Omega_{xyz}$ and $\Delta_p$, we use (4) and calculate the analytical expression for the upper-level population considering the condition that the probe field is very weak as compared to the driving fields and $\Omega_m = 0$, as

$$\rho_{11}(\Delta_p) = \frac{\Omega_p^2(\Delta_p - \Delta)^2}{[((\Delta_p - \Delta)\Delta_p - \Omega_{xyz}^2)^2 + \gamma_{21}^2(\Delta_p - \Delta)^2]}, \quad (5)$$

with $\Gamma_{32} = 0$ and $\gamma_{21} = \gamma_{31}$.

**Results and discussion –** We consider $\rho_{11}(\Delta_p)$ as the filter function which determines the 3D conditional position probability distribution of the atom. It is well known that for the $\Lambda$-type atom-field configuration the coherent population trapping occurs for the two photon resonance condition and the ground level population never goes to zero [24]. We, however, consider that the two detuning parameters $\Delta$ and $\Delta_p$ are far detuned from each other and thus avoiding the two photon resonance condition, as a result we can obtain maximum population of the excited level $|1\rangle$. For $\Delta = 0$, the probe detun-ing condition becomes $\Delta_p = 2\Omega(\sin kx + \sin ky + \sin kz)$ when $\xi = \varsigma = \phi = 0$ and $k_i = k$ ($i = 1,2,3,4,5,6$) which may lead to the maxima in the 3D conditional position probability distribution. One can notice that the position probability distribution of the atom in 3D space, which is conditioned upon the measurement of upper-level population or probe-absorption and leads to the 3D spatial measurement of the atom, is directly related to the probe field detuning. However, we consider a general case and carry out the analysis of the 3D atom localization numerically, i.e., without considering any approximation on $\Omega_p$ and $\Omega_i (i = 1,2,3)$.

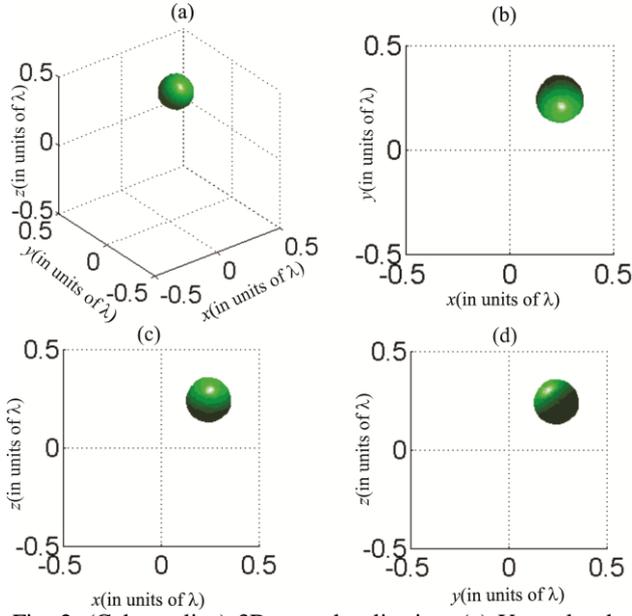

Fig. 2. (Color online) 3D atom localization: (a) Upper-level population as a function of κx, κy and κz, (b)-(d) xy, xz and yz view of (a), respectively. The other parameters are $\Omega = 2.3\Gamma$, $\Omega_p = 0.1\Gamma$, $\Omega_m = 0$, $\Delta = 0$, $\Delta_p = 13.8\Gamma$, $k_1 = k_3 = k_5 = 0.8\kappa$, $\eta = \varsigma = \varphi = \pi/8$, $\Gamma_{12} = \Gamma_{13} = 0.5\Gamma$ and $\Gamma_{32} = 0$ ($\Gamma = 6$MHz).

We first ignore the effect of Doppler shift and show the best possible spatial measurement of a single atom using Λ-type atom-field configuration. We select our parameters as $\Omega$ = 2.3Γ, $\Omega_p$ = 0.1Γ, $\Omega_m$ = 0 ($E_m$ = 0), $\Delta$ = 0, $\Delta_p$ = 13.8Γ, $k_1 = k_3 = k_5 = 0.8\kappa$, $\eta = \varsigma = \varphi = \pi/8$, $\Gamma_{12} = \Gamma_{13} = 0.5\Gamma$ and $\Gamma_{32} = 0$ ($\Gamma = 6$MHz). The values for the decay rate Γ and Rabi frequencies correspond to [23]. Following the same procedure as has been utilized in [7,16,20], we consider slightly different wavelengths and the phase shifts associated with the standing-wave fields. The plot of the filter function $\rho_{11}(\Delta_p)$ versus normalized positions κx, κy and κz shows a single 3D peak which is reflected as a sphere in the xyz space, see Fig. 2(a). To determine where exactly the center of the sphere is located, we show the 2D views of the filter function $\rho_{11}(\Delta_p)$ in Fig. 2(b-d). From these figures, we can observe that the sphere is centered at π/2, π/2, π/2 which determines the position of the atom in the 3D xyz space.

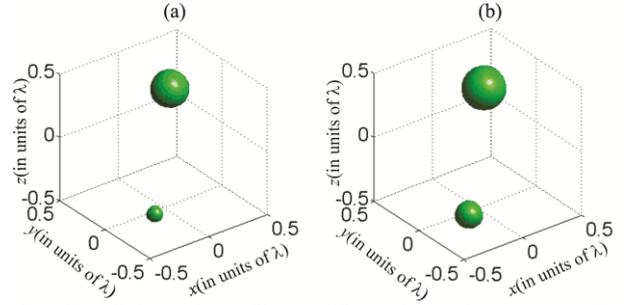

Fig. 3. (Color online) Plot of 3D conditional position probability distribution versus normalized positions $\kappa x$, $\kappa y$ and $\kappa z$ for different values of Doppler width $D$. (a) $D = 0.3\Gamma$ and (b) $D = \Gamma$. The remaining parameters are same as those used in Fig. 2(a).

Here we would like to emphasize that this is not the only possible position of the in the xyz space, one can get other positions as well. For example, by just changing the values of the phase shifts from $\eta = \varsigma = \pi/8$ to $\eta = \varsigma = -\pi/8$, one can a new position of the atom, i.e., $-\pi/2, -\pi/2, -\pi/2$, instead of $\pi/2, \pi/2, \pi/2$ in the xyz space. These results also show that the precision in spatial measurement of the single atom within a cubic wavelength is enhanced by a factor of 8 as compared with [21].

Until now, we have not incorporated effect of the Doppler shift; however, our main objective is to consider that the atom is not stationary during its interaction with the laser fields. and consider that the atom moves with velocity 'v' along the direction of the probe field [25]. In this case, the effect of Doppler broadening on the 3D spatial measurement of the atom must be considered. In the following, we consider that atom moves in the $z$ direction, i.e., perpendicular to the standing-wave fields acting along $x$ and $y$ directions. The fields only in the direction of motion of the atom are important to take into account the effect of Doppler shift. In the present case, only the probe field and one of the standing-wave fields, e.g., $E_z$ are considered propagating along the direction of the motion of the atom and we only consider these two fields to investigate the effect of Doppler broadening. As standing-wave field $E_z$ is constructed from two counter-propagating beams therefore for the resonant interaction between the atom and standing-wave field, there is no shifts in the frequency of the standing-wave field. Reason being, the counter-propagating fields cancel the shift of each other. Let us only consider the Doppler shift in probe field and replace the probe field detuning $\Delta_p$ with $\Delta_p + k_p v$ in Eq. (4), where $k_p$ is the wave-vector of the probe field and 'v' is the velocity of the atom. To obtain upper level population in a Doppler-broadened medium, velocity dependent $\rho_{11}(\Delta_p + k_p v)$ should be integrated over considering Maxwell-Boltzmann distribution for the atomic velocities. In the presence of Doppler effect, the upper-level population can be written as

$$\rho_{11}(d) = \frac{1}{\sqrt{2\pi}D} \int_{-\infty}^{\infty} \rho_{11}(\Delta_p + k_p v) e^{-(v)^2/2D^2} d(v), \quad (6)$$

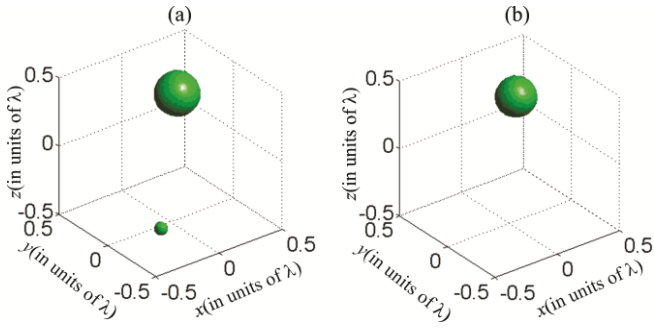

Fig. 4. Plot of 3D conditional position probability distribution versus normalized positions $\kappa x$, $\kappa y$ and $\kappa z$ for different values of microwave field $\Omega_m$. (a) $\Omega_m = 0.02\Gamma$ and (b) $\Omega_m = 0.03\Gamma$. The remaining parameters are the same as those used in Fig. 3(b).

where $D = \sqrt{k_p^2 K_B T / M}$ is Doppler width; while $K_B$, $T$ and $M$ are the Boltzmann constant, absolute temperature and mass of the atom, respectively. We call $\rho_{11}(d)$ as the Doppler broadened filter function for the 3D conditional position probability distribution.

In Fig. 3, we plot the Doppler broadened 3D filter function versus normalized positions κx, κy and κz in the xyz space for two different values of Doppler width, i.e., $D$ is equal to (a) $0.3\Gamma$ and (b) $\Gamma$ while keeping the other parameters same as in Fig. 2(a). The precision in position of the atom reduces as we introduce the Doppler effect and a second sphere starts appearing. The size of the second sphere which was not present in the absence of the Doppler broadening increases as we increase the value of the Doppler width, See Fig. 3(a) and (b). This shows that the precision in position measurement of the single atom reduces significantly in the presence of the Doppler shift. This implies that for a better spatial measurement of the atom having the best possible spatial resolution, one need to have a Doppler free spectroscopy which seems almost impossible until and unless one must have ultra-cold atoms. Another possibility one can think about is to eliminate the Doppler shift if the probe field is acting along x or y direction perpendicular to the motion of the atom or if one must consider that the motion of the atom is classical. The former, i.e., using ultra-cold atoms limits the practical applications, e.g., in atom lithography at room temperature. The latter is a more approximate solution and does not mimic the real scenario. In the following, we address this issue and discuss that in the presence of a microwave field, the effect of the Doppler shift can be reduced and the precision in spatial measurement of the single atom can be attained even in the presence of Doppler broadening.

We consider that a microwave field $E_m$ is applied which drives the atomic transition $|2\rangle \leftrightarrow |3\rangle$. For two different choices of the corresponding Rabi frequency $\Omega_m$, i.e., is equal to (a) $0.02\Gamma$ and (b) $0.03\Gamma$, we plot the Doppler broadened filter function $\rho_{11}(d)$ versus normalized positions κx, κy and κz. We notice that due to the presence of microwave field which couples the atomic transition $|2\rangle \leftrightarrow |3\rangle$ and generates atomic coherence between the levels $|2\rangle$ and $|3\rangle$ one can obtain precision in spatial measurement even in the presence of Doppler shift.

**Conclusion–** In conclusion, we have considered an experimentally realizable scheme based on Δ-type atom-field configuration [26] for the precision position measurement of the hot $^{87}$Rb atoms. Our scheme suggests a possible solution to overcome the issues in precise position measurement of the atom which is influenced by the Doppler shift. A single position of the atom, which is Doppler shifted, with high precision in 3D space can be obtained via control of an external microwave field.

***

Rahmatullah would like to thank Ziauddin for valuable discussions and Ministry of Science and Technology, Taiwan for the support